\documentclass{article}

\usepackage{amsmath} 

\usepackage{PRIMEarxiv}
\usepackage[T1]{fontenc}
\usepackage{lmodern} 
\usepackage[utf8]{inputenc} 
\usepackage{hyperref}       
\usepackage{url}            
\usepackage{booktabs}       
\usepackage{amsfonts}       
\usepackage{nicefrac}       
\usepackage{microtype}      
\usepackage{lipsum}
\usepackage{fancyhdr}       
\usepackage{graphicx}       
\graphicspath{{media/}}     

\title{VQSynergy: Robust Drug Synergy Prediction with Vector Quantization Mechanism
}

\author{
  Jiawei Wu*\\
  School of Data Science and Engineering \\
  East China Normal University \\
  Shanghai, China\\
   \And
  Mingyuan Yan \thanks{* equal contribution}\\
  School of Data Science and Engineering \\
  East China Normal University \\
  Shanghai, China\\
   \And
  Dianbo Liu \\
  School of Medicine and College of Design\&Engineering \\
  National University of Singapore \\
  Singapore\\
}

\begin{document}

\maketitle

\begin{abstract}
The advancement in optimizing cancer therapies has been significantly boosted by the ability to accurately predict drug synergies. While traditional approaches like clinical trials offer reliability, they are hampered by considerable time and financial requirements. The advent of high-throughput screening and computational breakthroughs has marked a shift towards more efficient methods for investigating drug interactions. In this study, we introduce VQSynergy, an innovative machine learning framework that leverages the Vector Quantization (VQ) mechanism, combined with gated residuals and a custom attention mechanism, to improve the accuracygeneralizability of drug synergy predictions. Our results show that VQSynergy outperforms existing models in robustness, especially in conditions with Gaussian noise, underscoring its superior efficacy and value in the intricate and often unpredictable field of drug synergy research. This study highlights VQSynergy's potential to transform the field through its advanced predictive capabilities, thus contributing to the refinement of cancer treatment strategies.
\end{abstract}


\section{Introduction}

\begin{figure}[hb!]
\centering 
\includegraphics[width=0.8\textwidth]{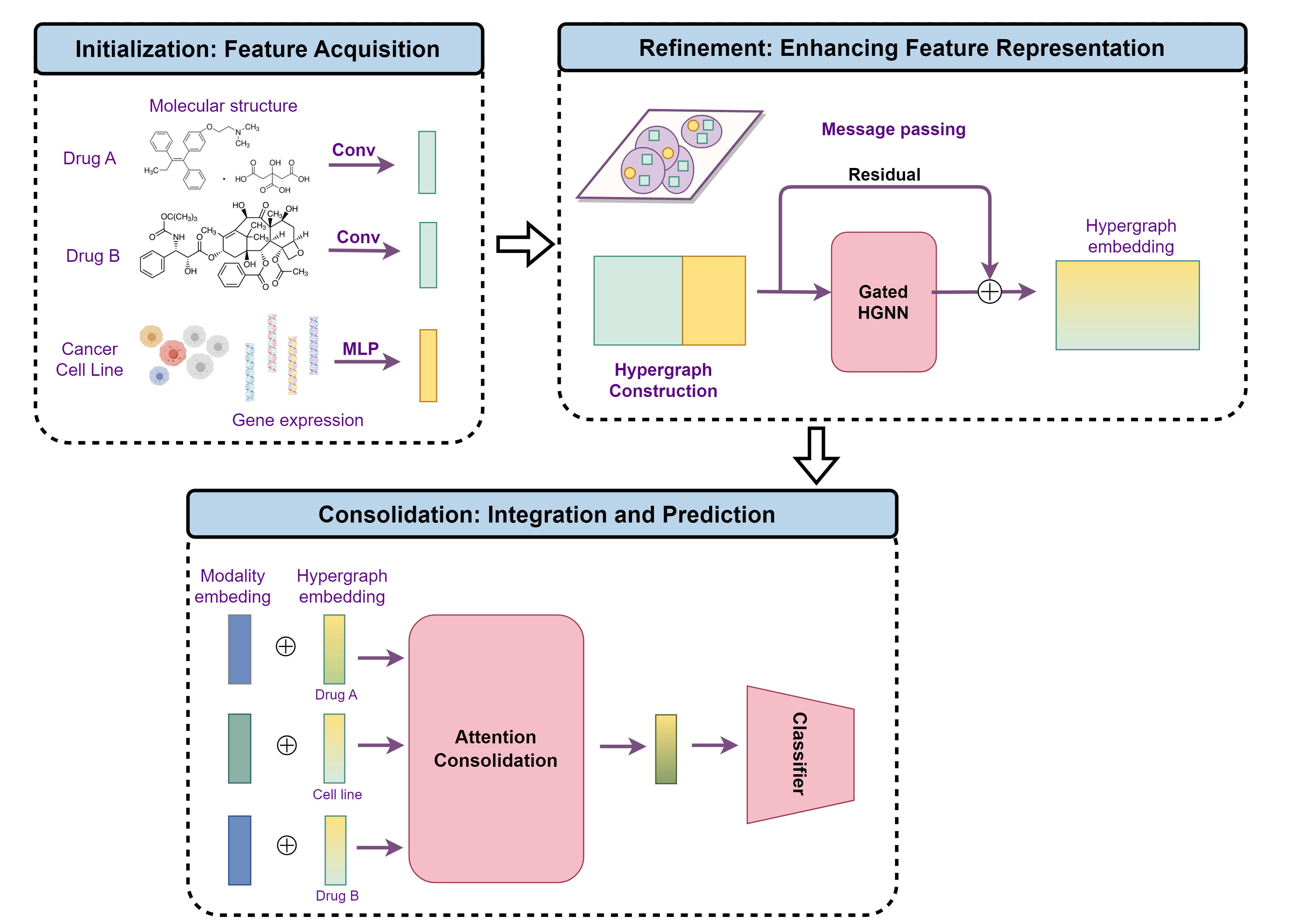} 
\caption{Overview of VQSynergy} 
\label{fig:pipeline} 
\end{figure}

The treatment of cancer faces challenges due to its complex nature. While the traditional `one gene, one drug' approach proposed by Ehrlich \cite{fitzgerald2006systems, hopkins2008network} remains common for many cancer types, it targets actively proliferating cells indiscriminately, causing damage to healthy tissue alongside cancerous cells. This oversimplified view of cancer biology overlooks the intricate interplay of factors involved. Further, monotherapy treatment is prone to drug resistance as continuous exposure to a single compound prompts cancer cells to activate alternative survival pathways \cite{gottesman2002multidrug, khdair2010nanoparticle, holohan2013cancer}. In response to these limitations, there has been a shift towards combination therapy, where drug synergy offers the potential for increased efficacy at lower doses, thereby reducing toxicity and the likelihood of drug resistance \cite{al2012combinatorial, mokhtari2013combination}. This paradigm shift emphasizes the need to reconsider our approach to cancer treatment, moving beyond the simplistic `one tumor, one gene, one drug' strategy towards a more holistic understanding of cancer biology \cite{sun2013high}.

The landscape of drug synergy prediction has been revolutionized by advances in high-throughput screening methods and computational modeling \cite{he2018methods}. Despite these advancements, the scarcity of synergistic drug pairs and the vast search space pose challenges, necessitating the development of computational models to prioritize viable combinations \cite{he2018methods}. These technologies have led to the construction of extensive datasets like O’Neil \cite{o2016unbiased}, NCI-ALMANAC \cite{holbeck2017national}, and others, capturing over a million experimentally measured synergy scores. These resources also introduce the complexity of handling vast amounts of data, underscoring the need for efficient computational models.

In addressing this need, recent advancements in machine learning and deep learning methods have shown promise in predicting drug synergy. DeepSynergy \cite{preuer2018deepsynergy} uses deep learning approach for predicting drug synergy, utilizing chemical and genomics data to outperform traditional machine learning methods in identifying novel synergistic drug combinations. The DTF \cite{sun2020dtf} model combines tensor factorization with a deep neural network (DNN) to predict drug synergy. Furthermore, HypergraphSynergy \cite{liu2022multi} predicts anti-cancer drug synergy by formulating synergistic drug combinations over cancer cell lines as a hypergraph, which often grapples with the intricate multi-way relations existing between drugs and cell lines. This study underscores the necessity to employ models that can capture and interpret these multi-dimensional relationships, leading to the development of hypergraph neural networks (HGNNs) \cite{feng2019hypergraph}, which have shown promise in various fields, including biomedical applications.

However, the message passing in HGNNs can hinder the model's generalization, especially with new node features absent in the training data. After message passing, these new features might inadvertently degrade predictive performance. To address this, a more meticulous approach to message passing is required. Implementing Vector Quantization (VQ) \cite{van2017neural} in the message passing process presents a promising solution. By conceptualizing message passing as a form of communication between different entities \cite{liu2021discrete}, the application of a simpler `lingua franca' can facilitate more efficient and effective interactions within the hypergraph framework.

Previous studies in drug synergy prediction have been significantly hindered by a heavy reliance on the quality of data pertaining to drug-associated biological entities. This dependency has often limited their performance, particularly in scenarios involving novel drugs or diseases where prior information is sparse. Recognizing this limitation, we developed \textbf{VQSynergy}, a robust framework for drug synergy prediction, employing a Vector Quantization (VQ) mechanism and gated residuals, coupled with a specially designed attention mechanism, VQSynergy not only achieves high predictive accuracy, but also enhances the model's generalization. This innovative approach ensures that our model is not constrained by the availability of prior knowledge, making it particularly effective for the early screening stages of drug development.

\section{Methods}
\subsection{Overview of VQSynergy}

Our methodology reconceptualizes drug synergy prediction within an advanced hypergraph framework, treating it as a hyperedge prediction problem. This innovative approach unfolds through three interconnected phases \ref{fig:pipeline}: \textbf{initialization}, \textbf{refinement}, and \textbf{consolidation}. Each phase plays a pivotal role in processing and integrating varied data types, thereby synthesizing comprehensive information crucial for accurate drug synergy predictions. Besides, our model introduces novel methodologies and integrates state-of-the-art techniques to substantially enhance the prediction accuracy and reliability in each phase of the analysis.

\subsection{Initialization: Feature Acquisition}
\label{sec:init}

The \textbf{initialization} stage is essential in our approach to predicting synergy involving collecting initial features for drugs and cell lines, and aligning them to a uniform dimensionality, which is critical for later analysis in our hypergraph model.

\textbf{Molecular Structure Data for Drugs}: For each drug, we construct molecular graphs \( \mathbf{G} = (\mathbf{X}_\text{atoms}, \mathbf{A}) \) using the Simplified Molecular Input Line Entry System (SMILES) \cite{kim2019pubchem}. In these graphs, the attribute matrix \( \mathbf{X}_\text{atoms} \), with dimensions \( \mathbb{R}^{n \times F_d} \), captures atomic attributes, and the adjacency matrix \( \mathbf{A} \), sized \( \mathbb{R}^{n \times n} \), represents atomic connections. To fit our model's requirements for one-dimensional data representations, we convert these multidimensional matrices into a singular feature vector \( \mathbf{x_d}\) for each drug, encapsulating its molecular structure. This conversion relies primarily on pooling techniques. Before pooling, we apply TransformerConv \cite{shi2020masked} to more effectively extract structural details from the molecular graphs.

\textbf{Genomic Data for Cell Lines}: For cell lines, we refine gene expression profiles into feature vectors \( \mathbf{x_c} \) through a process of normalization and transformation. This step is pivotal in capturing the distinct genomic characteristics of each cell line. To ensure compatibility with our model’s uniform dimensional requirements, we utilize a Multi-Layer Perceptron (MLP). The MLP functions to project these genomic features into a dimensionally consistent space. This projection is essential for integrating the genomic features of cell lines with the molecular structure data of drugs in the hypergraph model, ensuring a cohesive and comprehensive dataset for analysis.

\subsection{Refinement: Enhancing Feature Representation}
\label{sec:refine}

In the \textbf{refinement} phase of our model, we transform raw data into enhanced feature representations by constructing a drug synergy hypergraph. This process involves utilizing Hypergraph Neural Network Convolution, integrating gated residual connections, and employing Vector Quantization (VQ) techniques. The combination of these methods, particularly the inclusion of VQ, constitutes a meticulous message passing approach. This approach is crucial for accurately identifying complex interactions between drugs and cell lines, which are fundamental to effective drug synergy prediction. The incorporation of VQ allows for more precise feature discretization, further refining the model's ability to understand and predict intricate interrelationships in drug synergy applications. This phase, with its focus on detail and precision, significantly enhances the model's analytical capabilities, ensuring robust and reliable predictions in the complex field of drug synergy.

\subsubsection{Representing Multi-way Relations}

In this phase, we create a hypergraph to represent synergistic drug-cell line combinations. This hypergraph is essential for understanding the complex interactions of drugs and their effects on various cell lines, aiding in assessing the potential efficacy of drug combinations.

Nodes \( v \) in the hypergraph, initialized as detailed in Section~\ref{sec:init}, are described by the feature matrix \( \mathbf{X} = \begin{bmatrix} \mathbf{X}_D \\ \mathbf{X}_C \end{bmatrix} \). Here, \( \mathbf{X}_D \) denotes molecular structure features of drugs, and \( \mathbf{X}_C \) indicates gene expression features of cell lines.

The hypergraph structure is captured through an incidence matrix \( \mathbf{H} \in \mathbb{R}^{|V| \times |E|} \), focusing on the Drug-Cell Line Synergy Relationship. This relationship is central to the hypergraph, as it reveals the potential synergistic effects of specific drugs on different cell lines, which is crucial for predicting the effectiveness of drug combinations.

\subsubsection{Meticulous Message Passing in Graph}

In this section, we present a novel method for refining features in Graph Neural Networks (GNNs), emphasizing meticulous message passing. Our approach integrates hypergraph convolution, quantization strategies, and gated residual connections, tailored for modeling object interactions and predicting their future states, while embedding relational information.

A major challenge in GNNs is ``over-smoothing'' \cite{li2018deeper}, a phenomenon where repeated convolutions cause node features to become overly similar and indistinct. To combat this, we incorporate a residual connection to preserve feature uniqueness. The node feature update at \( l^\text{th} \) layer is formulated as:

\begin{equation}
    \mathbf{X}^{(l+1)} = \mathbf{X}^{(l)} + \Delta \mathbf{X}^{(l)},
    \label{eq:X_update}
\end{equation}

where \( \Delta \mathbf{X}^{(l)} \) represents the feature update. This update is derived from message updates and gating functions, subsequently discretized into a neural communication format \cite{liu2021discrete}, drawing inspiration from VQ-VAE \cite{van2017neural}. For each layer, we utilize a discrete space vector \( e \in \mathbb{R}^{L \times m} \), denoted by \( e^{(l)} \), with \( L \) as the size of the discrete latent space and \( m \) representing the dimension of each latent vector \( e_j \). We segment each target vector into \( G \) parts, or discretization heads, quantize each head individually using the layer-specific codebook, and concatenate the outcomes:

\begin{equation}
    \Delta \mathbf{X}^{(l)} = q^{(l)}(\mathbf{m}^{(l)} \odot \mathbf{i}^{(l)}, L, G),
    \label{eq:delta_X}
\end{equation}

To manage multi-way relationships characteristic of hypergraphs, where one hyperedge can connect multiple nodes, our model employs Hypergraph Neural Networks (HGNN) \cite{feng2019hypergraph, bai2021hypergraph}. The message passing in HGNN is described as:

\begin{align}
& \mathbf{m}^{(l)} = 
    \text{ReLU}(
    \mathbf{D}^{-1} \mathbf{H} \mathbf{B}^{-1} \mathbf{H}^{\top} 
    \mathbf{X}^{(l)} \Theta_{m}^{(l)}), \label{eq:m} \\
& \nonumber \mathbf{D} = \sum_e Y_{ve}, \quad \mathbf{B} = \sum_v Y_{ve}. \label{eq:DB}
\end{align}

where \( \mathbf{D} \) and \( \mathbf{B} \) denote diagonal matrices representing node and hyperedge degrees, respectively.

A key element of our model is its detailed and carefully designed message passing mechanism. This mechanism notably incorporates a gating function to modulate the influence of adjacent nodes. We strategically initialize the biases \( b \) with negative values. This initial setting enables the gating function to start functioning close to an identity operation, which is crucial for the efficient integration of features.

For the \( l^\text{th} \) layer, the gating function is defined as follows:

\begin{equation}
    \mathbf{i}^{(l)} = \sigma(\mathbf{X}^{(l)} \Theta_{i}^{(l)}),
    \label{eq:il}
\end{equation}

where \( \sigma \) is the sigmoid function, crucial for preserving the uniqueness of node features and , thereby contributing significantly to the model's learning dynamics.

The overall graph \textbf{refinement} update rule in our model is formulated as:

\begin{equation}
    \mathbf{X}^{(l+1)} = \mathbf{X}^{(l)} + 
    q^{(l)}(
    \text{ReLU}(
    \mathbf{D}^{-1} \mathbf{H} \mathbf{B}^{-1} \mathbf{H}^{\top} 
    \mathbf{X}^{(l)} \Theta_{m}^{(l)}) 
    \odot 
    \sigma(\mathbf{X}^{(l)} \Theta_{i}^{(l)}), L, G).
    \label{eq:your_equation}
\end{equation}

This meticulous message passing approach is instrumental in ensuring the effectiveness of our model. By accurately capturing and refining complex features, it significantly enhances the model's ability to make precise predictions in the context of drug synergy.

\subsection{Consolidation: Integration and Prediction}

In the culminating phase of our methodology, known as the \textbf{consolidation} phase, we synthesize the previously refined features into predictive insights. This is achieved through a binary classification model that leverages the enriched features from the hypergraph. It distinguishes between synergistic and non-synergistic drug-cell line interactions, providing a binary outcome for each drug synergy scenario. This phase is crucial for accurately identifying effective drug combinations.

\subsubsection{Predictive Model}

Our predictive model processes data formatted as tuples \((\mathbf{x}_{d_1}, \mathbf{x}_{d_2}, ..., \mathbf{x}_{d_n}, \mathbf{x}_{c})\), where \(\mathbf{x}_{d_i}\) represents a drug and \(\mathbf{x}_{c}\) a cell line. Each \(\mathbf{x}\) is an output from the \(l^{th}\) layer of a prior Meticulous Message Passing (see Section \ref{sec:refine}). A key feature of our model is its capability to interpret these tuples irrespective of the sequence of drugs, as their order is arbitrary and should not influence the outcome.

To accommodate the unordered nature of the drugs, we utilizes an attention mechanism, as outlined by Vaswani \cite{vaswani2017attention}. This mechanism, prior to position encoding, inherently does not assume any specific order among inputs, aligning well with our model's requirements. To further enhance the model’s expressiveness, we introduce learnable modality embeddings for both drugs (\(\mathbf{m}_{d}\)) and cell lines (\(\mathbf{m}_c\)). In our attention-based model, we define the components for each attention head \(j\) as query (\(\mathbf{Q}_j\)), key (\(\mathbf{K}_j\)), and value (\(\mathbf{V}_j\)):

\begin{equation}
    \mathbf{Q}_j = (\mathbf{x}_{c} + \mathbf{m}_c)^T W_Q^j,
    \label{eq:Q}
\end{equation}

\begin{equation}
    \mathbf{K}_j = \begin{bmatrix} 
        (\mathbf{x}_{d_1} + \mathbf{m}_d)^T W_K^j \\ 
        (\mathbf{x}_{d_2} + \mathbf{m}_d)^T W_K^j \\
        \vdots \\
        (\mathbf{x}_{d_n} + \mathbf{m}_d)^T W_K^j \\
        (\mathbf{x}_{c} + \mathbf{m}_c)^T W_K^j \\
    \end{bmatrix},
    \label{eq:K}
\end{equation}

\begin{equation}
    \mathbf{V}_j = \begin{bmatrix} 
        (\mathbf{x}_{d_1} + \mathbf{m}_d)^T W_V^j \\ 
        (\mathbf{x}_{d_2} + \mathbf{m}_d)^T W_V^j \\
        \vdots \\
        (\mathbf{x}_{d_n} + \mathbf{m}_d)^T W_V^j \\
        (\mathbf{x}_{c} + \mathbf{m}_c)^T W_V^j \\
    \end{bmatrix},
    \label{eq:V}
\end{equation}

where \(W_Q^j\), \(W_K^j\), and \(W_V^j\) are respective weight matrices for the query, key, and value vectors in the \(j^{th}\) attention head.

The computation in each attention head is executed as follows:

\begin{align}
    \mathbf{head}_j &= \mathrm{Attention}(\mathbf{Q}_j, \mathbf{K}_j, \mathbf{V}_j) \nonumber \\
    &= \mathrm{Softmax} \left(
        \frac
            {\mathbf{Q}_j \mathbf{K}_j^T}
            {\sqrt{d_k}}
    \right) \mathbf{V}_j,
    \label{eq:attention}
\end{align}

Following this, the model forms a comprehensive representation by concatenating the outputs from all attention heads:

\begin{equation}
    \mathbf{h}_{d_1, d_2, \ldots, d_n, c} = \mathrm{Concat}(\mathbf{head}_1, \mathbf{head}_2, \ldots, \mathbf{head}_h)W^O,
    \label{eq:hd}
\end{equation}

Ultimately, a Multi-layer Perceptron (MLP) is applied to this concatenated output to generate the final predictive output. The MLP processes the representation and delivers a scalar value, which is normalized to a range between 0 and 1 using a sigmoid activation function. This normalized output represents the probability of potential synergistic effects among the drugs:

\begin{equation}
    \mathbf{s}_{d_1, d_2, \cdots, d_n, c} = \text{MLP}(\mathbf{h}_{d_1, d_2, \cdots, d_n, c}),
    \label{eq:s}
\end{equation}

This structure enables our model to adeptly discerns intricate relationships between drugs and cell lines, without being influenced by the order in which the drugs are presented.

\subsubsection{Training objective}

Given the binary nature of our prediction task, we employ the cross-entropy loss function, a standard and effective choice for binary classification models. The cross-entropy loss is defined as:

\begin{equation}
L_{\mathrm{cross-entropy}} = -\frac{1}{N} \sum_{i=1}^{N} [y_i \log(\hat{y}_i) + (1 - y_i) \log(1 - \hat{y}_i)],
\label{eq:cross-entropy}
\end{equation}

Here, \( y_i \) represents the actual label (synergistic or non-synergistic) of the \( i \)-th sample, and \( \hat{y}_i \) is the predicted probability generated by the model for that sample.


\section{Experiments and Results}


\subsection{Datasets}
We collected four categories of data for our research, which include drug synergy data, molecular information for drugs, genomic characteristics of cancer cell lines, and indications for drug therapy in diseases from multiple publicly available databases. Here are the specific details:

\begin{itemize}
    \item \textbf{Drug synergy datasets}
    
    We gathered data on the synergy of anti-cancer drugs from two prominent large-scale tumor screening datasets – the O'Neil dataset \cite{o2016unbiased} and the NCI-ALMANAC dataset \cite{holbeck2017national}. The O'Neil dataset comprises 23,062 samples involving 38 unique drugs and 39 distinct human cancer cell lines. Each sample measures Loewe synergy scores for two drugs in combination with a specific cell line. The NCI-ALMANAC dataset contains 304,549 samples, including ComboScores for 104 FDA-approved drugs in pairings across the NCI-60 cell line panel.
    
    \item \textbf{Drug molecular structures}
    
    Information on the simplified molecular input line entry system (SMILES) of drugs was obtained from the PubChem database \cite{kim2019pubchem}.
    
    \item \textbf {Gene expression in cancer cell lines}
    
    Data on gene expression in cancer cell lines were sourced from the Cell Lines Project within the COSMIC database \cite{forbes2015cosmic}. In this context, we specifically considered the expression values of 651 genes related to the COSMIC cancer gene census. These expression values were subjected to logarithmic (log2) transformation and z-score normalization.
\end{itemize}

We conducted data preprocessing on both the NCI-ALMANAC and O'Neil datasets, excluding cell lines lacking gene expression data and drugs without SMILES information. Following this refinement, the NCI-ALMANAC dataset comprises 74,139 measurement samples of ComboScores for 87 drugs across 55 cancer cell lines, while the O'Neil dataset encompasses 18,950 samples of Loewe synergy scores for 38 drugs and 39 cancer cell lines. Subsequently, we removed drugs from the 'drug-disease dataset' that were not present in the aforementioned drug synergy datasets. This led to the extraction of indications for 111 diseases corresponding to 111 drugs within the NCI-ALMANAC dataset and 222 diseases corresponding to 222 drugs within the O'Neil dataset. These datasets are employed in our research study.

\begin{table}[htbp]
\centering
\caption{Comparison of Different Datasets}
\label{tab:dataset_comparison}
\resizebox{0.60\columnwidth}{!}{
\begin{tabular}{@{}lccc@{}}
\hline
Dataset     & \#Drugs & \#Cell lines & \#Samples \\ 
\hline
NCI-ALMANAC & 87      & 55           & 75,493    \\
O'Neil      & 38      & 39           & 18,950    \\
\hline
\end{tabular}
}
\end{table}

\subsection{Baselines}

In this study, we conducted a comparative analysis of our approach with several contemporary drug synergy prediction techniques. Here is a brief overview of each of the baseline methods:

\begin{itemize}
    \item DeepSynergy \cite{preuer2018deepsynergy}: It utilizes a three-layer feedforward neural network to predict synergy scores, incorporating gene expression as cell line features and three types of chemical descriptors as drug features.

    \item DTF \cite{sun2020dtf}: DTF extracts latent features from the drug synergy matrix through tensor factorization and employs them to train a deep neural network model for predicting drug synergy.

    \item HypergraphSynergy \cite{liu2022multi}: This method formulates synergistic drug combinations across cancer cell lines as a hypergraph. In this hypergraph, drugs and cell lines are represented by nodes, while synergistic drug–drug–cell line triplets are represented by hyperedges. It leverages the biochemical features of drugs and cell lines as node attributes. Additionally, a hypergraph neural network is designed to learn drug and cell line embeddings from the hypergraph and predict drug synergy.
\end{itemize}


\subsection{Experiment Setup}

To assess the predictive performance of our method in comparison to the baseline models, we adopted a cross-validation (CV) approach. This involved randomly partitioning each dataset into a CV set comprising 90\% of the samples and an independent test set containing the remaining 10\%. Within the CV set, we considered three distinct practical scenarios:

1. \textbf{Random CV}: In this scenario, a five-fold CV was executed by randomly dividing the samples into five equal folds. The aim here was to rediscover established anti-cancer drug synergies.

2. \textbf{Stratified CV for Cell Lines}: We implemented a five-fold CV by randomly splitting the samples at the cell line level, ensuring that the test set exclusively contained previously unseen cell lines that were not part of the training set.

3. \textbf{Stratified CV for Drug Combinations}: Similar to the previous scenario, we conducted a five-fold CV by partitioning the samples at the level of drug combinations. This was done to ensure that the test set exclusively comprised previously unencountered drug combinations not present in the training set.

Additionally, we employed an independent test set to evaluate the performance of our model and the baseline models. All methods were trained on the entire CV set and then used for making predictions on the independent test set, facilitating an objective evaluation.

For classification tasks, we transformed synergy scores into binary outcomes using a threshold established in prior studies \cite{preuer2018deepsynergy,sun2020dtf}. Samples with synergy scores exceeding 30 were categorized as positive, while those below this threshold were considered negative. 



\subsection{Performance Comparison and Analysis}

For the sake of evaluating the performance of VQSynergy, we
compare it with baselines mentioned above. The comparative results of different methods for drug synergy prediction, as summarized in Table \ref{tab:method_comparison}, distinctly highlights the superiority of the VQSynergy model across various testing scenarios. In the Random (Rand.) scenario, VQSynergy achieves the highest performance with a score of 0.8566, which is a noticeable improvement over its closest competitor, HypergraphSynergy, at 0.8447. This represents a performance increase of approximately 1.41\%. In the Stratified Cell scenario, VQSynergy again leads with a score of 0.7808, surpassing HypergraphSynergy's 0.7752 by a narrower margin of 0.56\%. The most significant performance differential is observed in the Stratified DrugComb scenario, where VQSynergy's score of 0.8002 outperforms HypergraphSynergy's 0.7745 by 2.57\%.

The overall analysis indicates that VQSynergy consistently outperforms other methods like DFT, DeepSynergy, and HypergraphSynergy. This consistent lead across different evaluation metrics underscores the robustness and effectiveness of the VQSynergy model. It also suggests that the incorporation of the Vector Quantization mechanism provides a substantial advantage in accurately predicting drug synergies, especially in complex scenarios represented by Stratified DrugComb. The performance of VQSynergy in these various scenarios not only demonstrates its efficacy but also highlights its potential as a reliable tool in the field of drug synergy prediction, capable of handling diverse and complex datasets with higher accuracy compared to existing methods.

\begin{table}[htbp]
\centering
\caption{Comparison of Different Methods (NCI-ALMANAC)}
\label{tab:method_comparison}
\resizebox{0.70\columnwidth}{!}{
\begin{tabular}{@{}lccc@{}}
\hline
Method & Rand. & Strat. CellLine & Strat. DrugComb \\
\hline
DFT                   & 0.8238 & 0.7551 & 0.7502 \\
DeepSynergy           & 0.8350 & 0.7035 & 0.7739 \\
HypergraphSynergy     & 0.8447 & 0.7752 & 0.7745 \\
VQSynergy             & \textbf{0.8566} & \textbf{0.7808} & \textbf{0.8002} \\
\hline
\end{tabular}
}
\end{table}

\begin{table}[htbp]
\centering
\caption{Comparison of Different Methods (O'Neil)}
\label{tab:method_comparison}
\resizebox{0.70\columnwidth}{!}{
\begin{tabular}{@{}lccc@{}}
\hline
Method & Rand. & Strat. CellLine & Strat. DrugComb \\
\hline
DFT                   & 0.9138 & 0.8387 & 0.8424 \\
DeepSynergy           & 0.9060 & 0.8188 & 0.8583 \\
HypergraphSynergy     & 0.9230 & 0.8554 & 0.8621 \\
VQSynergy             & \textbf{0.9232} & \textbf{0.8611} & \textbf{0.8715} \\
\hline
\end{tabular}
}
\end{table}

\subsection{Robustness Validation}

In this part, our experiment is structured to compare the performance of VQSynergy under two main scenarios: without Gaussian noise and with Gaussian noise (standard deviation, \( \sigma \) = 0.1). We evaluate the model in three different settings. The aim is to assess the effectiveness of the VQ mechanism in maintaining the model's predictive accuracy under noisy conditions.

Initially, the model was tested without the introduction of Gaussian noise. In this setting, as shown in Table \ref{tab:without_gaussian_noise_results}, the performance of VQSynergy without the VQ mechanism is marginally better than when the VQ mechanism is incorporated. The difference in performance is minimal, suggesting that under normal conditions without external noise, the addition of the VQ mechanism does not significantly enhance the model's capability.

The robustness of the VQSynergy model becomes particularly evident with the introduction of Gaussian noise (\( \sigma \) = 0.1), as shown in Table \ref{tab:gaussian_noise_results}. When the VQ mechanism is employed, there is a noticeable improvement in performance across all three scenarios. Specifically, the use of VQ results in an increase of 2.52\% for Random (82.19\% vs. 79.67\%), 0.64\% for stratified cell line (75.90\% vs. 75.26\%), and 1.60\% for stratified drug combination (76.77\% vs. 75.17\%) compared to the model without VQ mechanism. These results clearly illustrate the effectiveness of the VQ mechanism in enhancing the model's performance, particularly under conditions with data imperfections like Gaussian noise.

\begin{table}[htbp]
\centering
\caption{Results w/o Gaussian Noise}
\label{tab:without_gaussian_noise_results}
\resizebox{0.70\columnwidth}{!}{
\begin{tabular}{@{}lccc@{}}
\hline
Method & Rand. & Strat. CellLine & Strat. DrugComb \\
\hline
VQSynergy (w/o VQ)                     & 0.8591 & 0.7836 & 0.8014 \\
VQSynergy (w/ VQ)                      & 0.8566 & 0.7808 & 0.8002 \\
\hline
\end{tabular}
}
\end{table}

\begin{table}[htbp]
\centering
\caption{Results in Gaussian Noise (\( \sigma \) = 0.1)}
\label{tab:gaussian_noise_results}
\resizebox{0.70\columnwidth}{!}{
\begin{tabular}{@{}lccc@{}}
\hline
Method & Rand. & Strat. CellLine & Strat. DrugComb \\
\hline
VQSynergy (w/o VQ)           & 0.7967 & 0.7526 & 0.7517 \\
VQSynergy (w/ VQ)            & \textbf{0.8219} & \textbf{0.7590} & \textbf{0.7677} \\
\hline
\end{tabular}
}
\end{table}

\section{Conclusion and Future Direction}

In this study, we introduce VQSynergy, a robust framework employing the Vector Quantization (VQ) mechanism for enhanced drug synergy prediction. This innovative model combines VQ with gated residuals and a bespoke attention mechanism, setting a new standard in accuracy and generalizability in predictive modeling. Notably, robustness tests reveal that VQSynergy maintains superior predictive performance even under Gaussian noise, a testament to its resilience. This demonstrates the distinct advantage of incorporating the VQ mechanism in handling complex, noisy datasets typical in drug synergy studies.

The emerging trend towards higher-order combination therapies, as evidenced by the shift from traditional dual-drug approaches to more complex multi-drug regimens, presents an exciting frontier for future research \cite{davies2019accelerating,properzi2018dolutegravir,gotwals2017prospects}. This evolution, marked by the approval and investigation of therapies involving three or more drugs in diseases like cancer, HIV, and tuberculosis, signifies a rich area for exploration \cite{lakhtakia2015historical}. The success of regimens like the five-drug R-CHOP therapy for Diffuse Large B-Cell Lymphoma underscores the potential of such approaches \cite{lakhtakia2015historical}. In line with this progression, our future work will attempt to extend the VQSynergy model to analyze and predict the efficacy of these complex multi-drug combinations. The burgeoning repository of dose–response data for numerous drug combinations will be instrumental in advancing this research, paving the way for more effective and tailored therapeutic strategies \cite{tekin2018prevalence}.

Besides, the conventional VQ methodology, while foundational, is somewhat rudimentary and not without potential drawbacks. Specifically, the traditional 'arg min' approach in VQ could introduce issues in optimization and representation. To address this, an enhancement inspired by recent advancements, such as the method proposed in \cite{ramesh2021zero}, can be considered. This enhancement transforms the distance calculation into a probability distribution function and incorporates entropy as part of the loss function in this setup. Furthermore, the straight-through \cite{bengio2013estimating} method commonly used in VQ has been gradually superseded by the Gumbel-Softmax \cite{jang2016categorical, maddison2016concrete} approach. However, even more advanced gradient estimation techniques, as suggested by recent research \cite{paulus2020rao,fan2022training,liu2023bridging}, remain unexplored and warrant further investigation. These methods can be attempt to offer improvements in optimization and representation fidelity, thus enhancing the overall performance of VQ-based models in various applications.


\newpage
\bibliographystyle{unsrt}  
\bibliography{references}

\end{document}